\documentclass[a4paper,11pt]{article}
\pdfoutput=1 % if your are submitting a pdflatex (i.e. if you have
\usepackage{jheppub} % for details on the use of the package, please
\usepackage[T1]{fontenc} % if needed

\title{\boldmath Dark Matter Halo Axions Shift Hadron Collider $W$ Mass}

\author[a,1]{N. Bray-Ali,\note{Corresponding author.}}

\affiliation[a]{Science Synergy,\\Los Angeles, CA 90045, USA}
	\emailAdd{nbrayali@gmail.com}
	
	\date{November 11, 2026}

\abstract{The 2026 measurement of the $W$ mass by CMS at the Large Hadron Collider disagrees sharply with the earlier and more precise measurement by CDF at the Tevatron, yet the CMS result seems to agree rather well with the standard model value. A similar fact pattern is identified for two other observables that have been the subject of recent intense interest: the strong coupling constant and the leading-order hadronic contribution to the muon magnetic dipole moment. Using a straightforward proper-time analysis of the leading-order hard gluon corrections to the $W$ boson production amplitude from quark-antiquark annihilation in high-energy hadron collisions, it is suggested that the CDF experiment has discovered a new phenomenon beyond the standard model and that the CMS measurement confirms this discovery: A coherent dynamic background axion field created by axions from the local dark matter halo of the galaxy couples to the hard virtual gluon within the luminous four-volume at the interaction point inside the collider detector and this axion-gluon coupling shifts the $W$ mass determination linearly in the product of the beam width and the bunch length.}  
	
\begin{document}
\maketitle
\flushbottom
\section{Introduction}
The high-precision measurements of the charged weak $W$ vector boson mass, $M_W\approx80.4~{\rm GeV}$, with the CMS experiment at the Large Hadron Collider (LHC) in 2026 \cite{cms2026} and with the CDF II detector at the Tevatron in 2022 \cite{cdf2022} disagree by $5.4\sigma$, more than five standard deviations. Further, the best present global analysis of electroweak data in the standard model of particle physics \cite{pdg2024,global-ew2018,global-ew2022} yields a $W$ mass value that, while agreeing with the CMS result, nevertheless differs from the CDF determination by more than $7\sigma$. Given the conventional $5\sigma$ threshold used for the discovery of new phenomena in particle physics, such as the 2012 Higgs boson discovery by ATLAS \cite{higgs-atlas} and CMS \cite{higgs-cms}, or the 1995 top quark discovery by CDF \cite{top-cdf} and D0 \cite{top-d0}, the discrepancy between the CDF and CMS results thus constitutes the discovery of a phenomenon beyond the standard model of particle physics.

Within the standard model, high-energy hadron collisions produce both the charged $W$ and neutral $Z$ weak vector bosons primarily through quark-antiquark annihilation using the weak nuclear force \cite{ellis_book}. However, the strong nuclear force also plays a role, not just in determining the momenta of the quark and antiquark within the colliding hadrons, described by the parton distribution functions \cite{nnpdf2017,cteq2021,pdf4lhc,mmht2014}, and in soft collinear processes \cite{ladinsky1994,landry2003}, but also, critically, in hard contributions that shape the high-momentum tail of the boson transverse-momentum $p_T$ spectrum \cite{ellis_book}. At the hadron collision energies used in the CDF measurements, for example, these hard contributions from the strong nuclear force contribute roughly one-third of the boson production cross-section \cite{ellis_book}.

The strong nuclear force acts differently in the local dark matter halo of the galaxy \cite{eilers2019,nfw1997} than in the quantum chromodynamic (QCD) vacuum described by the standard model of particle physics \cite{politzer1973,wilczek1973,politzer1974}. In the context of high-energy colliders, a sign of this difference is the persistent tension --- greater than $3\sigma$ since 2011 --- between the best present electron-positron collider-based measurement of the strong coupling constant, $\alpha_s(M_Z)$, at the energy-scale set by the $Z$ boson mass \cite{iain2011}, and the standard model value estimated with lattice QCD \cite{lattice-alphas}. A crucial hint in this direction is that the tension with the lattice QCD estimate goes away when comparing instead to the best present $\alpha_s(M_Z)$ measurements using hadron collider detectors \cite{atlas-alphas,cdf-alphas}, which had a significantly smaller luminous four-volume at their interaction points than the electron-positron collider detectors used in the earlier measurement \cite{hep-collider-params-now,hep-collider-params-petra}.

In 2024 an analogous fact pattern emerged --- this time with significance exceeding the conventional $5\sigma$ discovery threshold --- with the measurement by the CMD-3 detector at the VEPP-2000 electron-positron collider \cite{cmd2024-long,cmd2023-short} of the leading order hadronic contribution to the muon magnetic dipole moment, $a_{\mu}^{\rm HLO-\rho}$, from the neutral $\rho^0$ vector meson that dominates photon-hadron interactions at low energy \cite{sakurai1963,sakurai1969,feynman1971,muon-theory2025}. Although the CMD-3 measurement agrees with the best present lattice QCD estimate \cite{bmw2021}, the single most precise such measurement, that of the KLOE detector at the DA$\Phi$NE electron-positron collider \cite{kloe2009,kloe2011,kloe2013}, falls more than $5\sigma$ below not just the best present standard model estimate \cite{bmw2024} but also the CMD-3 experiment \cite{cmd2024-long}. Crucially, the luminous four-volume at the interaction point within the KLOE detector \cite{kloe-luminosity} was significantly larger than that within CMD-3 \cite{cmd3-beams-current,vepp-2000}.

Indeed, the luminous four-volume created by the overlap in space and in time of the hadron bunches at the beam intersection point within the CDF detector at the Tevatron was significantly larger than that within CMS at the LHC \cite{hep-collider-params-now}. Following the fact pattern suggested by the measurements of $\alpha_s(M_Z)$ and $a_{\mu}^{\rm HLO-\rho}$ at high-energy colliders, it is then perhaps understandable not only that the value of $M_W$ found at CDF significantly differs from the value found at CMS, but also that the standard model $M_W$ value agrees with the result found by the detector, CMS, with the smaller luminous four-volume. Still, to make the case compelling, we must find the link between the size of the luminous four-volume within high-energy collider detectors and the size of effects that shift strong interactions in high-energy collisions away from standard model expectations.

The link is that QCD axions \cite{wilczek1978,weinberg1978} from the local dark matter halo of the galaxy \cite{axion-dark-matter-theory,axion-dark-matter-experiment} with rest-mass, $m_A$, and energy density, $\rho_A$, create a coherent dynamic background field with four-dimensional Fourier amplitude, $\tilde{\phi}_A(q_A)$, at the axion lab-frame four-momentum, $q_A\approx(m_Ac,0,0,0)$, whose components are all much smaller than those of the hadrons within the luminous four-volume, $VT$, probed by the high-energy collider experiment (SI units here and throughout):\footnote{While the size of the luminous four-volume, $VT$, drops out of properly normalized scattering amplitudes \cite{ellis_book}, the effect considered here is {\em not} a scattering amplitude. Instead, the coherent dynamic background axion field is delocalized over the entire luminous four-volume created at the interaction point within the detector by the overlap of the pair of colliding hadron bunches in space and in time, while the hard virtual gluon to which the coherent dynamic background axion field couples is localized on a scale set by the $W$ boson lab-frame momentum, $p_W$. See Eq.~(\ref{eq:gluon-green-shift}) for the effect of the axion-gluon coupling to this coherent dynamic background axion field on the gluon Green's function which, in turn, leads to the $W$ mass shift seen in hadron colliders. }
\begin{equation}
	\label{eq:axion-amp}
	\tilde{\phi}_A(q_A)=\frac{1}{m_Ac^2}\sqrt{\frac{\rho_AVT}{\hbar}}.
\end{equation} 
To be concrete, the axion energy density, $\rho_A=0.30~(3)~{\rm GeV/cm^3}$ \cite{eilers2019}, is given by the best present astrophysical estimate of the local dark matter halo energy density, while the beam intersection volume, $V=8\pi\sigma_x\sigma_y\sigma_z$ and the bunch crossing-time $T=2\sigma_z/c$, are set by the beam width, $\sigma_x$, the beam height, $\sigma_y$, and the bunch length, $\sigma_z$, at the interaction point within the collider detector (CMS at LHC Run 2, $\sigma_x=\sigma_y=8.5~{\mu\rm m}, \sigma_z=8.0~{\rm cm}$; CDF at Tevatron Run 2, average of $p$ and $\overline{p}$, $\sigma_x=\sigma_y=22~{\mu\rm m}, \sigma_z=48~{\rm cm}$ \cite{hep-collider-params-now}). Although the axion rest-mass, $m_A$, will turn out not to enter explicitly the size of the $W$ mass shift, $\Delta M_W$, caused by the axion background field, $\tilde{\phi}(q_A)$, nevertheless it still enters implicitly through the axion decay constant, $f_A$, which determines $m_A$ given the best present estimate of the topological susceptibility of the QCD vacuum, $\chi_{\rm QCD}$ \cite{wilczek1978, weinberg1978,gorghetto}:
\begin{equation}
	\label{eq:axion-mass-formula}
	m_Ac^2=\frac{\sqrt{\chi_{\rm QCD}}}{f_A}=0.569~(5)~{\rm eV}\left(\frac{10^{7}~{\rm GeV}}{f_A}\right).
\end{equation} 

Explicitly, the link between the $W$ mass shift, $\Delta M_W$, and the luminous four-volume, $VT$, takes the following simple form, given by the $W$ boson momentum in the lab-frame, $p_W\approx6.24~(5)~{\rm GeV/c}$ \cite{w-momentum}, and the weak boson mass ratio, $M_Z/M_W\approx1.13$ \cite{pdg2024}:
\begin{eqnarray}
	\label{eq:w-mass-shift-formula}
	\frac{\Delta M_W}{M_W}&=&\ln\left(\frac{M_Z}{M_W}\right)\frac{\Delta\alpha_s(M_W)}{\alpha_s(M_W)}\nonumber\\
	&=&\ln\left(\frac{M_Z}{M_W}\right)\tilde{\phi}_A(q_A)\left(m_Ac^2\right)\left(p_Wc\right)g_{Agg}\nonumber\\
	&=&\frac{\alpha_s(M_W)}{8\pi}\frac{p_wc}{f_A}\sqrt{\frac{\rho_AVT}{\hbar}}\ln\left(\frac{M_Z}{M_W}\right).
\end{eqnarray}  
Here, the shift $\Delta\alpha_s(M_W)$ in the strong coupling constant at the $W$ mass-scale, $\alpha_s(M_W)\approx0.120$ \cite{pdg2024}, is caused by the linear coupling with strength, $g_{Agg}=\alpha_s(M_W)/(8\pi f_A)$ \cite{weinberg1978,wilczek1978,axion-dark-matter-theory}, between the coherent dynamic axion background field, $\tilde{\phi}_A(q_A)$, and the dot product, ${\bf E}\cdot{\bf B}$, of the chromo-electric field ${\bf E}$ and the chromo-magnetic field ${\bf B}$ \cite{weinberg1978,wilczek1978,axion-dark-matter-theory}. For axion rest-mass energy, $m_Ac^2\approx0.5~{\rm eV}$, we then find the following straightforward resolution of the $5.4\sigma$ tension, $M_W({\rm CDF})-M_W({\rm CMS})=73~(14)~{\rm MeV}$ \cite{cms2026,cdf2022}, in terms of the relative shift caused by the difference in the luminous four-volumes probed by the two detectors:
	\begin{eqnarray}
		\label{eq:w-mass-shift-axion-mass}
		\Delta M_W({\rm CDF})&-&\Delta M_W({\rm CMS})\nonumber\\
		&=& 70~(4)~{\rm MeV/c^2} ~\left(\frac{m_Ac^2}{0.5~{\rm eV}}\right)		
	\end{eqnarray}

The paper is organized as follows. The proper-time method is used in section \ref{sec:method} to evaluate the shift of the leading-order quantum chromodynamic (QCD) contribution to the quark-antiquark annnihilation amplitude for charged weak $W$ vector boson production. Section \ref{sec:results} presents the resulting shift of the $W$ mass in hadron collider measurements. Section \ref{sec:discussion} discusses the significance of the $W$ mass shift in the context of recent precision hadronic spectroscopy. Finally, the concluding section \ref{sec:conclusions} summarizes the paper.
\section{Methods}
\label{sec:method}
 The proper-time method was introduced in 1951 by Schwinger to find the leading-order quantum electrodynamic (QED) contributions to the electron mass and magnetic dipole moment \cite{schwinger1951}. Here, it is used to evaluate the shift of the leading-order quantum chromodynamic (QCD) contribution to the quark-antiquark annnihilation amplitude for charged weak $W$ vector boson production \cite{politzer1974, ellis_book}. Hence, the starting point is the modified Dirac equation for the quark, $\psi(x)$, in interaction with its proper chromodynamic radiation field, and the external weak nuclear gauge field, $A_{\mu}(x)$ \cite{schwinger1951, politzer1974,ellis_book},
\begin{equation}
	\label{eq:quark-dirac}
	\gamma_{\mu}\left(-i\partial_{\mu}-gA_{\mu}(x)\right)\psi(x)+\int(dx')M(x,x')\psi(x')=0.
\end{equation}    
To the second order in the strong charge $g_s$, the mass operator, $M(x,x')$, is given by, 
\begin{equation}
	\label{eq:mass-operator}
	M(x,x')=m_0\delta(x-x')+ig_s^2\gamma_{\mu}G(x,x')\gamma_{\mu}D_+(x-x').
\end{equation}
%color factor?
Here, $G(x,x')$ is the Green's function of the Dirac equation in the external field, and $D_+(x-x')$ is a gluon's Green's function, expressed by,
\begin{equation}
	\label{eq:gluon-green}
	D_+(x-x')=\frac{1}{(4\pi)^2}\int_0^{\infty}\frac{dt}{t^{2}}\exp\left[i\frac{(x-x')^2}{4t}\right].
	\end{equation} 
	
Now, the coherent dynamic background field, $\tilde{\phi}_A(q_A)$, is created by axions from the local dark matter halo of the galaxy within the luminous four-volume probed in the collider detector during the $W$ mass measurement \cite{axion-dark-matter-theory,axion-dark-matter-experiment}. The fundamental axion-gluon coupling has Lagrangian density \cite{weinberg1978,wilczek1978,axion-dark-matter-theory}, 
\begin{equation}
	\label{eq:axion-gluon-lagrangian}
	\mathcal{L}_{ag}=g_{Agg}\phi_A(x){\bf E}\cdot{\bf B},
\end{equation}
where $\phi_A(x)$ is the four-dimensional inverse Fourier transform of the coherent dynamic background axion field, $\tilde{\phi}_A(q_A)$, and the axion-gluon coupling strength is $g_{Agg}=\alpha_s(M_W)/(8\pi f_A)$. To first order in $g_{Agg}$, the axion-gluon coupling yields a shift, $\Delta D_+(x-x')$, in the gluon Green's function, that grows linearly with the coherent dynamic background axion field, $\tilde{\phi}_A(q_A),$\footnote{See ref.~\cite{schwinger1951} for the directly analogous shift of the electron Green's function  in the presence of a coherent dynamic background electromagnetic field as a result of the electron-photon coupling. Just as the photon field couples linearly to the electron electromagnetic current which contains a product of a pair of electron operators, so also the axion field couples linearly to the two-gluon operator formed by the dot product, ${\bf E}\cdot{\bf B}$, of the chromo-electric field, ${\bf E}$, and the chromo-magnetic field, ${\bf B}$. Hence, the shift in the gluon Green's function, shown in Eq.~(\ref{eq:gluon-green-shift}), takes an analogous form to the shift of the electron Green's function derived in ref.~\cite{schwinger1951}. }
\begin{equation}
	\label{eq:gluon-green-shift}
	\Delta D_+(x-x')=g_{Agg}\tilde{\phi}_A(q_A)\left(m_Ac^2\right)\left(p_Wc\right)D_+(x-x'),
\end{equation}   	
where we have used the fact that the axion rest-mass energy, $m_Ac^2\approx0.5~{\rm eV}$, suggested by Eq.~(\ref{eq:w-mass-shift-axion-mass}), falls in the gap, $\hbar/T\ll m_Ac^2\ll p_Wc$, such that we can treat the axion dynamics as a coherent dynamic oscillation on the time-scale set by the bunch crossing-time $T$, yet, at the same time, we can approximate the axion four-momentum, $q_A\approx0$, when performing the four-momentum convolution of the axion amplitude, $\tilde{\phi}_A(q_A)$ and the four-dimensional Fourier transform of the virtual gluon propagator, $\tilde{D}_+(q)$, to solve the gluon equations of motion for the change, $\Delta D_+(x-x')$, at leading-order in the axion-photon coupling, $g_{Agg}$.

Returning to the calculation of the leading-order QCD contribution to $W$ boson production from quark-antiquark annihilation, the shift in the gluon Green's function, $\Delta D_+(x-x')$, is equivalent to keeping the QCD vacuum gluon Green's function, $D_+(x-x')$, and instead simply shifting by $\Delta\alpha_s(M_W)$ the value of the strong coupling constant,\footnote{At leading order in the strong coupling constant, $\alpha_s(M_W)$, the contribution of the hard virtual-gluon exchange between annihilating quark and antiquark to the production of the $W$ boson goes as the product, $\alpha_s(M_W)(D_+(x-x')+\Delta D_+(x-x'))$, of the strong coupling constant and the virtual gluon Green's function, $D_+(x-x')+\Delta D_+(x-x')$ \cite{ellis_book}. Clearly, it is equivalent to replace this with the expression, $(\alpha_s(M_W)+\Delta\alpha_s(M_W))D_+(x-x')$, where the shift of the strong coupling constant, $\Delta\alpha_s(M_W)$, is given by Eq.~(\ref{eq:shift-alphas}).} 
\begin{equation}
	\label{eq:shift-alphas}
	\frac{\Delta\alpha_s(M_W)}{\alpha_s(M_W)}=\frac{\Delta D_+(x-x')}{D_+(x-x')}=g_{Agg}\tilde{\phi}_A(q_A)\left(m_Ac^2\right)\left(p_Wc\right).
\end{equation}
Just as the leading-order QED contribution to the electron mass has a logarithmic dependence on the low-energy cut-off of the proper electrodynamic radiation field \cite{schwinger1951}, so also the leading-order QCD contribution to $W$ boson production picks up a dependence on the logarithm of $M_W$, which serves as the low-energy cut-off here \cite{politzer1974}. In practice, the $Z$ boson mass, $M_Z$, provides the high-energy cut-off of the logarithm due to the procedure for tuning the strong coupling constant used in hadron collider measurements of $M_W$ \cite{cms2026,cdf2022}, and we thus find the schematic form for the leading-order QCD contribution, $\mathcal{A}(u\overline{d}\rightarrow W^+)$, to the amplitude for $W$ boson production from quark-antiquark annihilation in hadron collisions \cite{ellis_book}, 
\begin{equation}
	\label{eq:amplitude}
	\mathcal{A}(u\overline{d}\rightarrow W^+)\propto\alpha_s(M_W)\ln\left(\frac{M_Z}{M_W}\right).
\end{equation}
  
  \section{Results} 
  \label{sec:results}
  Keeping fixed the amplitude, $\mathcal{A}(u\overline{d}\rightarrow W^+)$, for the leading-order QCD contribution to $W$ boson production from quark-antiquark annihilation in hadron collisions, the change in the strong coupling constant, $\Delta\alpha_s(M_W)$, due to the coherent dynamic background axion field, $\tilde{\phi}_A(q_A),$ then leads to the shift, $\Delta M_W$, in the value of the $W$ mass extracted from the amplitude,\footnote{Both the CDF and CMS collaborations calibrate the value of the strong coupling constant, {\em in situ}, using measurements at the $Z$ boson rest-mass energy \cite{cdf2022,cms2026}. The resulting value for the strong coupling constant used in the analysis of $W$ production is, thus, effectively pinned at the standard model value, $\alpha_s(M_W)$. To keep fixed the quality of the fit to the observed production amplitude, $\Delta\mathcal{A}(u\overline{d}\rightarrow W)=0$, the shift in the actual strong coupling constant, $\Delta\alpha_s(M_W)$, given in Eq.~(\ref{eq:shift-alphas}), requires the collaborations to then make a compensating shift, $\Delta M_W$, shown in Eq.~(\ref{eq:ashift}), for the value of $M_W$ determined by their analysis.}
  \begin{eqnarray}
  	\label{eq:ashift}
  	0&=&\Delta \mathcal{A}(u\overline{d}\rightarrow W^+)\nonumber\\
  	&=&\Delta\alpha_s(M_W)\ln\left(\frac{M_Z}{M_W}\right)-\alpha_s(M_W)\frac{\Delta M_W}{M_W}.
  \end{eqnarray}
  Solving for the $W$ mass shift, $\Delta M_W$, and using the expression in Eq.~(\ref{eq:shift-alphas}) for the strong coupling shift, $\Delta \alpha_s(M_W)$, in terms of the coherent dynamic background axion amplitude, $\tilde{\phi}_A(q_A)$, given in Eq.~(\ref{eq:axion-amp}), we find the result stated in the introduction in Eq.~(\ref{eq:w-mass-shift-formula}). The $W$ mass shifts by an amount that scales linearly with the background axion amplitude, $\tilde{\phi}_A(q_A)$, and thus grows linearly with product of the bunch length, $\sigma_z$ and the beam radius, $\sigma_r=\sqrt{\sigma_x\sigma_y}$: 
  \begin{equation}
  	\label{eq:bunch-beam}
  		\Delta M_W = 7.11~(36)~{\rm MeV}~\left(\frac{\sigma_r}{10~{\rm \mu m}}\right)\left(\frac{\sigma_z}{10~{\rm cm}}\right)\left(\frac{m_Ac^2}{0.5~{\rm eV}}\right).
  \end{equation} 
%Combine the p_T spectrum at O(\alpha_s) with the shift in \alpha_s to get shift in M_W. 
    
  As shown in the introduction, the differences in beam parameters during the $W$ mass measurements by CMS and CDF largely explain the $5.4\sigma$ discrepancy between their results, provided the axion rest-mass, $m_A\approx 0.5~{\rm eV/c^2}$, is roughly one million times less than the electron rest-mass. While we have essentially treated $m_A$ as a free parameter in the analysis of the CMS-CDF $W$ mass discrepancy, it should be noted that there exists a deep and wide literature attempting to predict, constrain, and even measure the rest-mass energy of QCD axion dark matter \cite{axion-dark-matter-theory,axion-dark-matter-experiment,brayali2026,metrax2026}. Although this is not the place for a systematic engagement with that literature --- in particular the attempt in 2025 to ``rule out'' the value $m_Ac^2\approx0.5~{\rm eV}$ using blank-sky observations with the James Webb Space Telescope \cite{pinetti2025,pinetti2025_prl,roy2025} --- nevertheless, an estimate for $m_Ac^2$ can be made by simply identifying the predicted $W$ mass shift, $\Delta M_W({\rm CDF})=74.3~(3.7) ~(m_Ac^2/0.5~{\rm eV})$, for the CDF beam parameters, with the observed $7.2\sigma$ tension, $M_W({\rm CDF})-M_W({\rm SM})=75.5~(10.5)~{\rm MeV}$ \cite{cdf2022,pdg2024},
  \begin{equation}
  	\label{eq:axion-mass-determination}
  	m_Ac^2 = 0.5~{\rm eV}\frac{75.5~(10.5)}{74.3~(3.7)} = 0.508~(75)~{\rm eV}.
  \end{equation} 
  
  By contrast, a result that is largely independent of the size of $m_A$ comes from considering the sign of the $W$ mass shift. Having fixed the sign of $\Delta M_W$ to match the observed shift, $M_W({\rm CDF})-M_W({\rm CMS})$, the sign can be predicted for the shift of the strong coupling constant, $\Delta\alpha_s(M_Z),$ seen in the measurements at electron-positron colliders mentioned in the introduction \cite{iain2011}. In these measurements, the strong interaction contributes to the hard exchange of large transverse-momentum between the quark and antiquark that are created by the decay of $Z$ bosons made in electron-positron annihilation.
  
  Remarkably, time-reversal symmetry, $\mathcal{T}$, relates the overall process, $e^+e^-\rightarrow Z\rightarrow q\overline{q},$ to electron-positron pair production, $q\overline{q}\rightarrow Z\rightarrow e^+e^-$, from quark-antiquark annihilation at hadron colliders \cite{ellis_book}. Given the $\mathcal{T}$ symmetry of QCD \cite{politzer1973,wilczek1973}, all of the machinery for calculating the leading-order QCD contribution carries over to the $e^+e^-$ collider case. However, the dot product, ${\bf E}\cdot{\bf B}$, of the chromo-electric and chromo-magnetic fields, flips sign under $\mathcal{T}$, just as the analogous product of electric and magnetic fields does \cite{weinberg1978,wilczek1978}.
  
 Thus, the change of the gluon propagator, $\Delta D_+(x,x'),$ must also flip sign under time-reversal, since the sign of the axion amplitude, $\tilde{\phi}_A(q_A)$, is fixed under the change from hadron to $e^+e^-$ collisions. But then, we find that the sign of the shift in the strong coupling constant, $\Delta \alpha_s(M_Z)<0$, must be negative in the $e^+e^-$ case. Remarkably, the observed tension shows this sign with the $e^+e^-$ determination, $\alpha_s(M_Z,e^+e^-)=0.1135~(11)$ \cite{iain2011}, falling $3.8\sigma$ {\em below} the lattice QCD estimate, $\alpha_s(M_Z,~{\rm QCD})=0.1183~(7)$ \cite{lattice-alphas}. 
  
Historically, the axion was first proposed nearly half a century ago as a way of killing the term, $\mathcal{L}_{\theta}=(g_{Agg}f_A)\theta_A{\bf E}\cdot{\bf B}$, where now, $\theta_A$, is just some real number constant of nature \cite{weinberg1978,wilczek1978}. Under time-reversal, the value of $\theta_A$ flips sign, and thus a non-zero value for $\theta_A$ would generically break $\mathcal{T}$. However, the logical consequence of having axions in the local dark matter halo, as we have shown, is a coherent dynamic background axion field, $\tilde{\phi}_A(q_A)$, that changes the way the strong nuclear force acts in a manner that transforms under the discrete space-time symmetries of QCD as the product ${\bf E}\cdot{\bf B}$: Odd under spatial inversion, $\mathcal{P}$, odd under time-reversal, $\mathcal{T}$, yet even under charge-conjugation, $\mathcal{C}$ \cite{weinberg1978,wilczek1978}.

  \section{Discussion} 
  \label{sec:discussion}Having discussed the size of the $W$ mass discrepancy and the sign of the $\alpha_s(M_Z)$ tension, the last case to discuss of recent intense interest is the hadronic leading order $\rho^0$ resonance contribution to the muon magnetic dipole moment, $a_{\mu}^{\rm HLO-\rho}$. As emphasized in the introduction, the fact pattern in this case traces a familiar line in which the recent measurement, by the CMD-3 collaboration, $a_{\mu}^{\rm HLO-\rho}({\rm CMD-3})=3793~(30)\times10^{-11}$ \cite{cmd2024-long}, lands close to the best available lattice QCD estimate \cite{bmw2021,muon-theory2025}. Meanwhile, the more established and precise measurement, by the KLOE collaboration, probed a much larger luminous four-volume and gives a $5.1\sigma$ different result, $a_{\mu}^{\rm HLO-\rho}({\rm KLOE})=3606~(21)\times10^{-11}$ \cite{kloe2009,kloe2011,kloe2013}, that also sharply disagrees -- by $5.6\sigma$ -- with the best present standard model value \cite{bmw2024}.  
  
  Now, the coherent dynamic background axion field, $\tilde{\phi}_A(q_A)$, couples the isovector neutral $\rho^0$ vector meson that dominates photon-hadron interactions at low-energy to its isoscalar cousin, the $\omega$ vector meson whose existence is required by the otherwise puzzling fact that the neutron, $n$, lacks electric charge despite having opposite isospin, $T_3(n)=-T_3(p)$, compared to the proton, $p$ \cite{sakurai1963, sakurai1969}:
  \begin{equation}
  	\label{eq:axion-rho-omega}
  	\mathcal{L}_{A\rho\omega}= g_{A\rho\omega}\phi_A(x)\left({\bf E}_{\rho}\cdot{\bf B}_{\omega}+{\bf B}_{\rho}\cdot{\bf E}_{\omega}\right).
  \end{equation}  
Here the hadro-electric field, ${\bf E}_{V}$, and hadro-magnetic field, ${\bf B}_{V}$, for $V=\rho^0,~\omega$, are given in terms of the hadro-dynamic field strength tensor, $F_{\mu\nu}=\partial_{\mu}V_{\nu}-\partial_{\nu}V_{\mu}$, in the usual manner \cite{sakurai1969}. Vector meson dominance determines the coupling strength, $g_{A\rho\omega}=878~g_{A\gamma\gamma}$ \cite{feynman1971}, in terms of the axion-photon coupling strength, $g_{A\gamma\gamma}\approx\alpha/(2\pi f_A)$ \cite{weinberg1978,wilczek1978,axion-dark-matter-theory,axion-dark-matter-experiment}, with $\alpha\approx1/137$ the electromagnetic ``fine-structure'' coupling constant in QED analogous to the strong coupling constant, $\alpha_s$, in QCD.
  
  This coupling between the coherent dynamic background axion field, $\tilde{\phi}_A(q_A)$, and the neutral $\rho^0$ vector mesons produced in $e^+e^-$ annihilation experiments, such as CMD-3 and KLOE, converts these isovector mesons into their isoscalar $\omega$ cousins. However, these experiments are designed to measure the $\rho^0$ resonance contribution, $a_{\mu}^{\rm HLO-\rho}$, through the reaction, $e^+e^-\rightarrow\rho^0\rightarrow\pi^+\pi^-$, in which the $\rho^0$ decays to a pair of oppositely charged pions \cite{muon-theory2025}. The isoscalar $\omega$ decays instead to three pions, $\omega\rightarrow\pi^+\pi^-\pi^0$, and it is treated as off-resonant background in the $a_{\mu}^{\rm HLO-\rho}$ measurement \cite{kloe2013}.
  
In this way, the sign of the $a_{\mu}^{\rm HLO-\rho}$ discrepancy can be understood. The detector with significantly larger luminous-volume at the interaction point, KLOE, converted more $\rho^0$ to $\omega$ as a result of its larger value for the coherent dynamic background axion field, $\tilde{\phi}_A(q_A)$. Hence, KLOE lands {\em low} for $a_{\mu}^{\rm HLO-\rho}$ compared to CMD-3 which was less efficient at converting $\rho^0$ to $\omega$, given its small luminous four-volume, and thus lands close to the standard model, in agreement with the observed sign of the $a_{\mu}^{\rm HLO-\rho}$ discrepancy.   
  
At the same qualitative level of discussion, it is straightforward to see that the {\em charged} $\rho^{\pm}$ vector mesons cannot couple to the axion in the same way as the neutral $\rho^0$ vector meson. Indeed, the $\omega$ vector meson is neutral, as is the axion, and thus simple electrical charge conservation forbids the coherent dynamic background axion field, $\tilde{\phi}_A(q_A)$, from messing with $e^+e^-$ collider-based determinations of $a_{\mu}^{{\rm HLO}-\rho}$ using the charged $\rho^{\pm}$ vector meson resonance, regardless of the size of the luminous four-volume probed in the interaction region of the collider detector. In striking agreement with this reasoning, the Muon $g-2$ Theory Initiative reported in 2020 the value for an $e^+e^-$ collider-based determination of the full hadronic leading order contribution, $a_{\mu}^{\rm HLO}(\rho^{\pm})=7030~(44)\times10^{-11}$ \cite{muon-theory2021,dhmz},\footnote{See ref.~\cite{muon-theory2021}, pg. 32, below Eq. (2.13).} that is dominated by measurements on the charged $\rho^{\pm}$ resonance \cite{belle,aleph}, and which {\em does not} significantly differ from the best present lattice QCD determination, $a_{\mu}^{\rm HLO}({\rm QCD})=7075~(55)\times10^{-11}$ \cite{bmw2021}.  
  
In passing, this discussion motivates a rather straightforward yet nevertheless rigorous and precise estimate of the standard model value for the full hadronic leading-order contribution, $a_{\mu}^{\rm HLO}$, based on three largely independent, mutually consistent, and comparably precise determinations. In addition to the charged-$\rho^{\pm}$ based determination, $a_{\mu}^{\rm HLO}(\rho^{\pm})=7030~(44)\times10^{-11}$ \cite{muon-theory2021,dhmz}, and the lattice QCD result, $a_{\mu}^{\rm HLO}({\rm QCD})=7075~(55)\times10^{-11}$ \cite{bmw2021}, the third determination simply builds on the CMD-3 result for the resonant $\rho^0$ contribution, $a_{\mu}^{\rm HLO-\rho}({\rm CMD-3})=3793~(30)\times10^{-11}$ \cite{cmd2024-long}, by tacking on the off-$\rho$ contributions, $a_{\mu}^{\rm HLO-off\rho}=3225~(29)\times10^{-11}$ \cite{muon-theory2025}, to yield the result, $a_{\mu}^{\rm HLO}(\rho^0)=7018~(42)\times10^{-11}$. Given the independence, precision, and consistency of these three determinations, we simply combine them in a weighted average to produce the standard model estimate,
\begin{equation}
	\label{eq:smamu-hlo}
	a_{\mu}^{\rm HLO}=7036~(27)\times10^{-11}.
\end{equation}
  
  Continuing this thread, the discussion then leads to a straightforward yet precise and rigorous estimate of the present tension between the standard model muon magnetic dipole moment, $a_{\mu}({\rm SM})$, and the best present determination using spin precession of high-energy muons in a magnetic storage ring, $a_{\mu}({\rm SR})$ \cite{fnal-final}. Simply tacking on the off-HLO contributions, $a_{\mu}^{\rm off-HLO}=116~584~879~(16)\times 10^{-11}$ \cite{muon-theory2021}, along with the estimate of the HLO contribution, $a_{\mu}^{\rm HLO}=7036~(27)\times10^{-11}$, gives a rigorous value for the standard model muon magnetic dipole moment with comparable precision to the storage ring value $a_{\mu}({\rm SR})$ \cite{fnal-final}:
  \begin{eqnarray}
  	\label{eq:smamu}
  	a_{\mu}({\rm SR})&=&116~592~070.5~(14.5)\times10^{-11}\nonumber \\
  	a_{\mu}({\rm SM})&=& 116~591~915~(31)\times10^{-11}.
  \end{eqnarray}
  Taking the difference yields a $4.6\sigma$ tension, $a_{\mu}({\rm SR})-a_{\mu}({\rm SM})=156~(34) \times10^{-11}$, that falls just short of the conventional threshold for discovery of a new phenomenon.

  \section{Conclusions}
  \label{sec:conclusions}
   The 2022 CDF high-precision measurement of the $W$ mass revealed the existence of a phenomenon in particle physics lying beyond the standard model. The 2026 CMS experiment result confirmed its existence and offered a critical clue to its nature: It goes away when you squeeze the luminous four-volume at the interaction point within the particle detector of a high-energy collider. The phenomenon can be understood in terms of the coherent dynamic background axion field created by the energy density of axions from the local dark matter halo of the galaxy within the luminous four-volume.
  
  Crucial to the analysis was the identification of a similar fact pattern to the $W$ mass discrepancy in two other cases involving precision measurement with high-energy colliders: the strong coupling constant, $\alpha_s(M_Z)$, and the $\rho^0$ vector meson contribution to the the muon magnetic dipole moment, $a_{\mu}^{\rm HLO-\rho}$. In each case, an older more precise measurement, analogous to CDF in the $W$ mass case, probed a larger luminous four-volume and landed far from the standard model. Meanwhile, a younger less precise measurement, analogous to CMS, probed a smaller luminous four-volume and landed near the standard model.
  
  Pursuing this analogy, the strong nuclear force acts differently in the local dark matter halo of the galaxy than in the quantum chromodynamic (QCD) vacuum described by the standard model. In the $W$ mass case, this difference was captured by a simple shift of $M_W$ going as the product of the beam width and the bunch length. For $\alpha_s(M_Z)$, the sign of the shift observed with electron-positron colliders was predicted based on the sign of the $W$ mass shift observed with hadron collisions. 
  
  Turning to $a_{\mu}^{\rm HLO-\rho}$, a consensus of independent and precise determinations yields a rigorous and straightforward estimate of the standard model value. The consensus relies, crucially, on the analogy with the $W$ mass and $\alpha_s(M_Z)$ cases to explain why the single most precise determination of $a_{\mu}^{\rm HLO-\rho}$, by the KLOE experiment, falls below the standard model value. The consensus was then used to show that a sharp tension exists --- just short of the conventional discovery threshold --- between the standard model muon magnetic dipole moment and the value observed with accelerated muon beams in magnetic storage rings.
  
 From the size of the $W$ mass discrepancy between CDF and the standard model, the mass of the axion was estimated to be roughly one million times less than the electron, around 0.5 eV. The estimate was based on a straightforward analysis of the leading-order hard QCD contribution to $W$ boson production from quark-antiquark annihilation using the proper-time method introduced by Schwinger. The key step was to shift the virtual gluon propagator that enters this hard contribution by using the fundamental axion-gluon coupling and the coherent dynamic background axion field created within the luminous four-volume by axions from the local dark matter halo of the galaxy: The $W$ mass shift seen by CDF came not from the production of axion dark matter, as one might suppose, but rather from its destruction by virtual gluons created in the hadron collision that coupled to this coherent dynamic axion background field.

\end{document}